\documentclass{elsart4-1}

\usepackage{amssymb}

\usepackage{graphics}
\usepackage{graphicx}
\usepackage{epsfig}
\usepackage{amssymb}
\journal{Physica B}

\begin{document}

\begin{frontmatter}


 \title{Transition between SU(4) and SU(2) Kondo effect}

\author[label1]{L. Tosi} 
\author[label2]{P. Roura-Bas}
\author[label1]{A. A. Aligia}
\address[label1]{Centro At\'omico Bariloche and Instituto Balseiro, Comisi\'on Nacional de
Energ\'{\i}a At\'omica, 8400 Bariloche, Argentina}
\address[label2]{Dpto de F\'{\i}sica, Centro At\'{o}mico Constituyentes, Comisi\'{o}n 
Nacional de Energ\'{\i}a At\'{o}mica, Buenos Aires, Argentina}
 \corauth[cor1]{Tel: 54-11 67727102; FAX: 54-11 67727121; roura@tandar.cnea.gov.ar}





\begin{abstract}
Motivated by experiments in nanoscopic systems, we study a generalized Anderson, which consists of two
spin degenerate doublets hybridized to a singlet by promotion of an electron to two conduction bands,
as a function of the energy separation  $\delta$ between both doublets. For $\delta$=0 or very large, 
the model is equivalent
to a one-level SU(N) Anderson model, with N=4 and 2 respectively.
We study the evolution of the spectral density for both doublets ($\rho_{1 \sigma}(\omega)$ 
and $\rho_{2 \sigma}(\omega)$) and their width in the Kondo limit
as $\delta$ is varied,
using the non-crossing approximation (NCA). As $\delta$ increases, the peak at the Fermi energy 
in the spectral density (Kondo peak) splits and the density of the doublet of higher energy $\rho_{2 \sigma}(\omega)$
shifts above the Ferrmi energy.
The Kondo temperature $T_K$ (determined by the half width at half maximum of the Kondo peak in 
density of the doublet of lower energy $\rho_{1 \sigma}(\omega)$)
decreases dramatically. The variation of $T_K$ with $\delta$ is reproduced by a simple variational calculation.  

\end{abstract}

\end{frontmatter}


\section{Introduction}

The Kondo effect, found originally for systems with magnetic impurities in
metals is now present in a variety of nanoscopic systems, including
semiconducting quantum dots \cite{kou}, magnetic adatoms on surfaces \cite{kou,lobos,trimer} 
and carbon nanotubes \cite{nyg}. In the latter, in
addition to the spin Kramers degeneracy, there is in addition orbital
degeneracy due to the ``pseudospin'' degree of freedom related with 
the particular band structure of graphene. This leads to an SU(4) Kondo
effect which has been observed experimentally \cite{jari,maka} and also
discussed theoretically \cite{lim,buss,and}. In particular, Lim \textit{et al.}
have studied the spectral density when the SU(4) symmetry is reduced to
SU(2), mainly by a change in the tunneling matrix elements \cite{lim,and}.

Our main motivation in the problem arises from interference phenomena in
systems of quantum dots \cite{scs,fea,soc} or molecules \cite{fea,bege,mole}. For
example depressions in the integrated conductance through a ring 
described by the Hubbard or $t-J$ model,
pierced by an Aharonov-Bohm magnetic flux, related with spin-charge separation 
\cite{scs,fea}, are due to a partial destructive interference when the energy of
two doublets cross \cite{fea}. Similar interference effects were predicted in
molecular transistors \cite{bege,mole}.
The effective Hamiltonian near the crossing
is discussed in the next section. 

To our knowlege, the conductance in these systems has so far
been calculated using approximate expressions or a slave-boson formalism \cite{soc}, 
which is valid only for very low temperatures and applied bias
voltages. This work is a step towards a more quantitative theory to describe
the transport through similar systems, treating the effective Hamiltonian
within the non-crossing approximation (NCA) \cite{nca,nca2}. Work is in
progress to deal with the non-equilibrium situation, which is necessary
within our formalism to calculate the current.

In this paper, we report on our study of the spectral density of the model as a function of the
splitting $\delta $ between both doublets in the Kondo regime. We also calculate the
dependence of the Kondo temperature $T_{K}$ with $\delta $ and analyze the
validity of the Friedel sum rule \cite{fri}.

\section{Model}

We start from a model in which two doublets of an interacting system are
hybridized with a singlet by promotion of an electron to two conducting
leads. This is the low-energy effective Hamiltonian for several systems with
partial destructive interference, such as Aharonov-Bohm rings \cite{fea}
or aromatic molecules \cite{bege,mole} connected to
conducting leads. The  Hamiltonian can be written as \cite{fea}

\begin{eqnarray}
H &=&E_{s}|0\rangle \langle 0|+\sum_{i\sigma }E_{i}|i\sigma \rangle \langle
i\sigma |+\sum_{\nu k\sigma }\epsilon _{\nu k}c_{\nu k\sigma }^{\dagger
}c_{\nu k\sigma }  \nonumber \\
&&+\sum_{i\nu k\sigma }(V_{i\nu }|i\sigma \rangle \langle 0|c_{\nu k\sigma }+{\rm H.c}.),  
\label{ham}
\end{eqnarray}%
where the singlet $|0\rangle $ and the two doublets $|i\sigma \rangle $ ($%
i=1,2$; $\sigma =\uparrow $ or $\downarrow $) denote the localized states 
(representing for example the low-energy states of an isolated molecule), 
$c_{\nu k\sigma }^{\dagger }$ create conduction states in the left
($\nu =L$) or right ($\nu =R$) lead, and $V_{i\nu }$ describe the four hopping
elements between the two leads and both doublets, assumed independent of $k$.

Changing the phase of the conduction states and the relative phase between
both doublets, three among the four  $V_{i\nu }$ can be made real and
positive. The phase $\phi $ of the remaining hopping  $|V|e^{i\phi }$,
depends on the particular system and its symmetry. For example in molecules
with rotational symmetry $\phi =(K_{1}-K_{2})l$, where $l$ is the distance between the 
sites connected to the left and right leads, and $K_{i}$ is the wave
vector of the state $|i\sigma \rangle$, which can be modified with an applied
magnetic flux \cite{fea}. In absence of magnetic flux and when the relevant
states have wave vector $K_{i}=\pm \pi /2$, as in rings with a number of
atoms multiple of four,  $\phi =\pi $ and there is complete destructive
interference in transport \cite{fea,mole}. 

In the following we will assume
this case, with states 1 and 2 related by symmetry implying   $|V_{1\nu
}|=|V_{2\nu }|$. We \ further assume symmetric leads, $\epsilon
_{Lk}=\epsilon _{Rk}=\epsilon _{k}$, $|V_{iL}|=|V_{iR}|$. Then, without loss
of generality we can take $V_{1L}=V_{1R}=V_{2L}=V>0$, $V_{2R}=-V$, and $E_2 \geq E_1$. 

For these parameters,
changing basis $c_{1k\sigma }^{\dagger }=(c_{Lk\sigma }^{\dagger
}+c_{Rk\sigma }^{\dagger })/\sqrt{2}$, $c_{2k\sigma }^{\dagger
}=(c_{Lk\sigma }^{\dagger }-c_{Rk\sigma }^{\dagger })/\sqrt{2}$, the
Hamiltonian takes the form of an SU(4) Anderson model with a "field" $\delta
=E_{2}-E_{1}$ and on-site hybridization $V^{\prime }=\sqrt{2}V$

\begin{eqnarray}
H &=&E_{s}|0\rangle \langle 0|+\sum_{i\sigma }E_{i}|i\sigma \rangle \langle
i\sigma |+\sum_{ik\sigma }\epsilon _{k}c_{ik\sigma }^{\dagger }c_{ik\sigma }
\nonumber \\
&&+V^{\prime }\sum_{ik\sigma }(|i\sigma \rangle \langle 0|c_{ik\sigma }+{\rm H.c}.).  
\label{ham2}
\end{eqnarray}

Interchanging the doublet index (1 or 2) with the spin index $\sigma$, one realizes 
that this model also 
describes transport trough a carbon nanotube with electron density depleated
at two points (so as to created an SU(4) quantum dot in the middle) under a real applied
magnetic field.

\section{Spectral density}

\begin{figure}
\includegraphics[width=7.5cm]{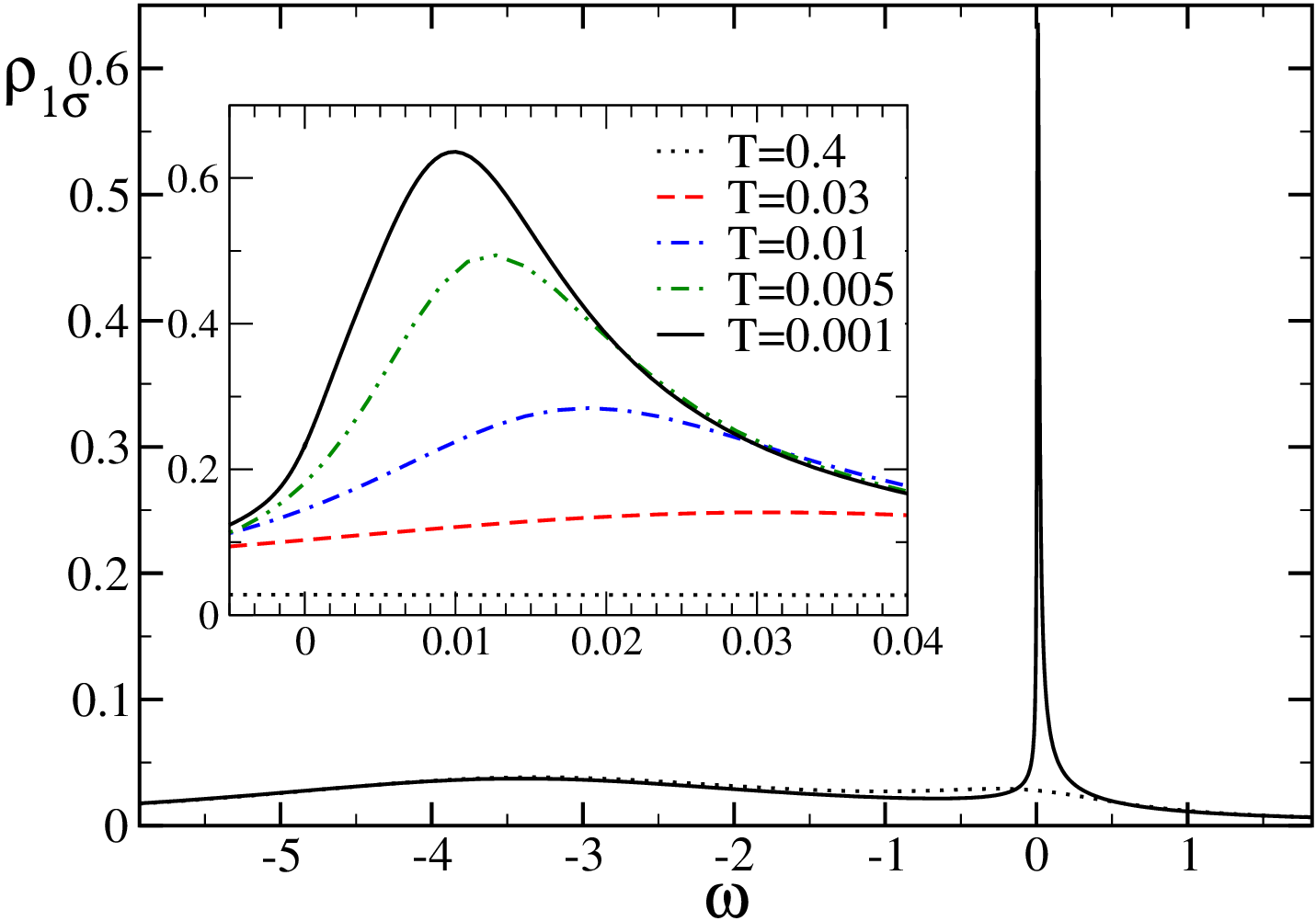}
\caption{Spectral density as a function of energy for different temperatures. The inset shows a detail
near the Fermi energy. Parameters are $\Delta =0.5$. $D=10$, $E_1=E_2=-4$. The lowest temperature is
$T= 10^{-3}=0.076T_K$. }
\label{dens1}
\end{figure}

In Fig. \ref{dens1}  we present numerical results for the spectral density $\rho _{i\sigma }$ of
the SU(4) Anderson model ($E_1=E_2$) in the Kondo regime 
$\epsilon_{F}-E_{i}\gg \Delta$, where  the hybridization function 
$\Delta=\pi \sum_{k}(V^{\prime })^{2}\delta (\omega -\epsilon _{k})$, assumed
independent of energy.  We set $\Gamma=2\Delta =1$ as the unit of energy. We also assume a
conduction band symmetric around the Fermi level $\epsilon _{F}=0,$ of half
width $D=10$. 

The spectral density shows a broad charge transfer peak near $E_1$. 
For temperatures below a characteristic energy scale $T_K$ (defined below), 
$\rho _{i\sigma }$ develops 
a narrow peak around the Fermi level. In contrast to the better known one-level SU(2) 
case, this peak is displaced towards positive energies and is much broader,
as discussed in Section 5.


\begin{figure}
\includegraphics[width=7.5cm]{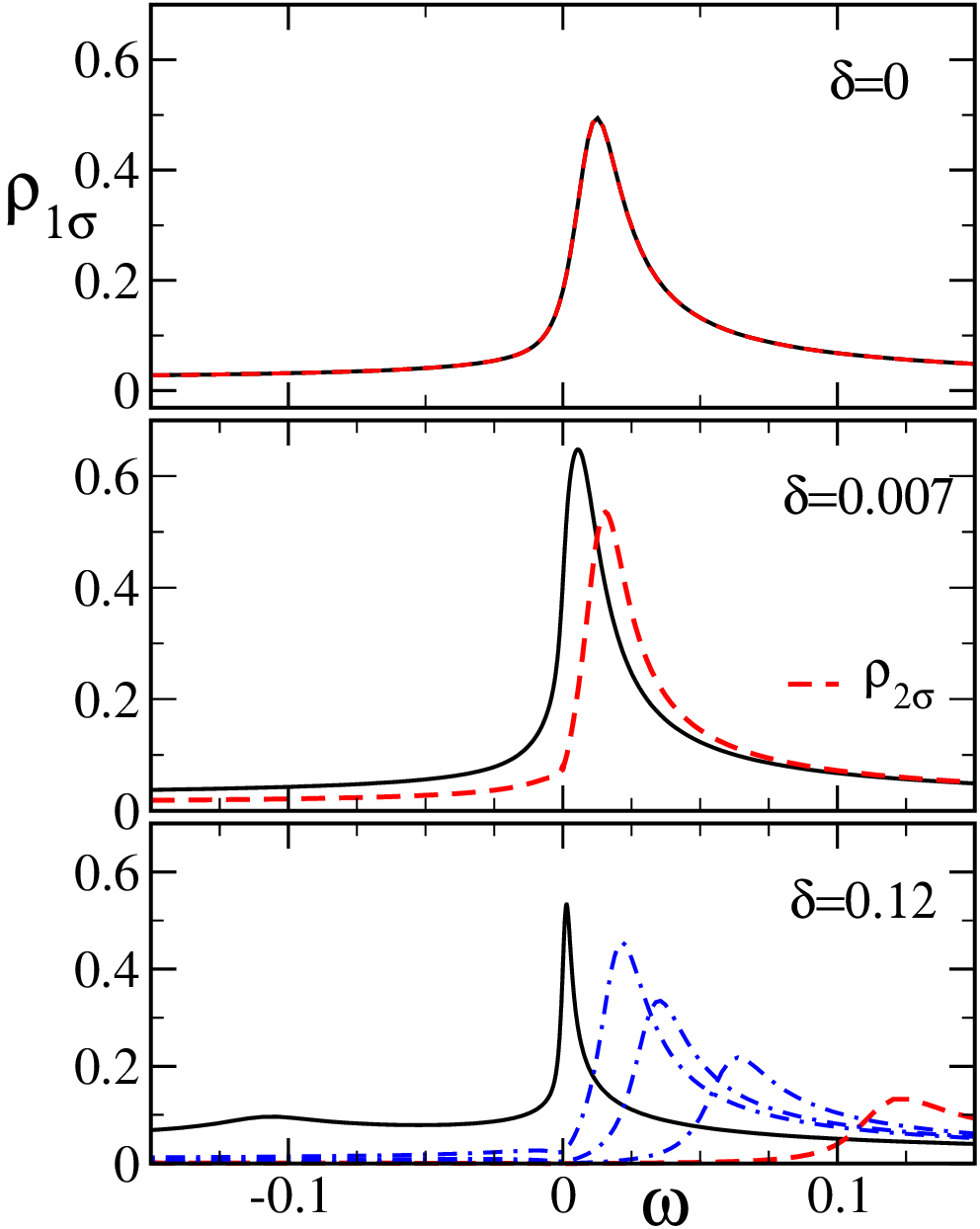}
\caption{Spectral densities for levels 1 (full line) and 2 (dashed line) 
as a function of energy for
different values of $E_2=E_1+\delta$ and $T= 10^{-3}$. Other parameters as in Fig. 1.
Dot-dashed line corresponds to $\rho _{2 \sigma }$ for $\delta=0.015$, 0.3 and 0.6.}
\label{dens}
\end{figure}

The evolution of the spectral densities as $E_2$ is displaced to larger energies, 
breaking the SU(4) symmetry is shown in Fig. \ref{dens}.  
The peak near the Fermi energy of $\rho _{2 \sigma} (\omega)$ is displaced towards positive energies
(near $\delta=E_2-E_1$). In contrast, the corresponding peak in $\rho _{1 \sigma } (\omega)$ narrows
significantly and displaces towards the Fermi energy. This implies that the Kondo temperature $T_K$
defined as the half width at half maximum of this peak, also decreases strongly.
The evolution of $T_K$ with $\delta$ is discussed in the Section 5.

In addition $\rho _{1 \sigma }$ develops a broad peak near energy $-\delta$ which becomes visible
when $\delta$ becomes greater than $T_K$.

\section{Friedel sum rule}

The Anderson model studied has a Fermi liquid ground state which satisfies well known relationships at
zero temperature. 
One of them is the Friedel sum rule which relates the spectral density at the Fermi level for each 
``pseudospin'' channel with the occupation of that channel \cite{fri}. For the simplest case of a constant density 
of conduction states, this rule reads 

\begin{equation}
\rho _{i\sigma }(\epsilon_{F})=\frac{1}{\pi \Delta }\sin ^{2}(\pi n_{i\sigma }),
\label{fsr}
\end{equation}%
where $n_{i\sigma }=\langle |i\sigma \rangle \langle i\sigma |\rangle $.

This is an exact relationship for a Fermi liquid, which is not necessarily satisfied by approximations.
In particular, it is known that at very low temperatures, the NCA has a tendency to develop
spurious spikes in $\rho _{i\sigma }(\omega)$ at the Fermi energy, while thermodynamic properties, 
such as expectation values are accurately reproduced \cite{nca,nca2}. 

\begin{figure}
\includegraphics[width=7.5cm]{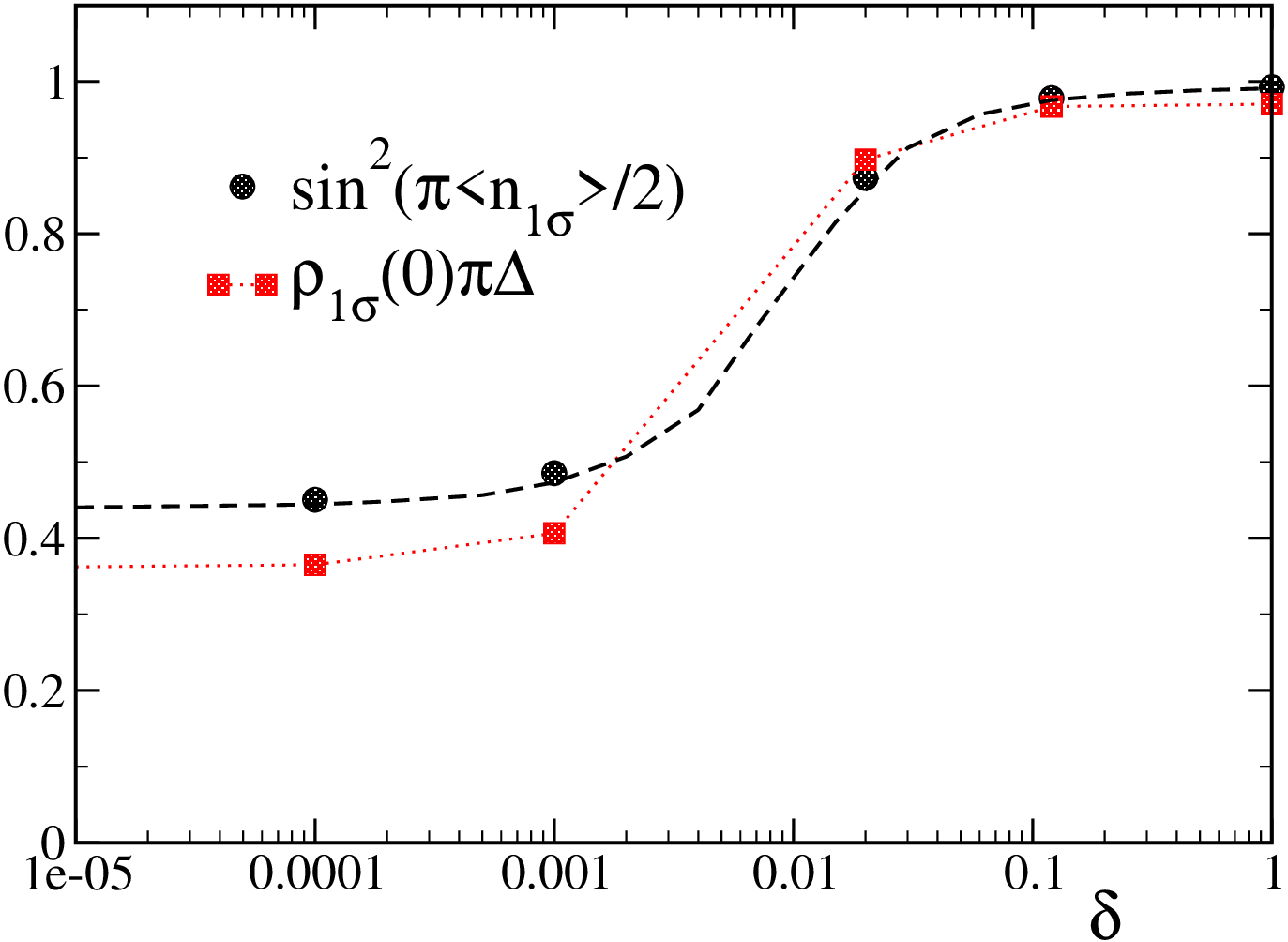}
\caption{Squares: rescaled spectral density of the lowest lying level at the Fermi energy as a 
function of $\delta$. Circles: corresponding (more accurate) result given by Eq. (\ref{fsr}). 
Lines are guides to the eye.
Parameters are  $T=0.1 T_K$ and the rest as in Fig. 1.}
\label{friedel}
\end{figure}

In Fig. \ref{friedel}, we compare both members of Eq. (\ref{fsr}) for the lowest lying doublet, 
at a temperature $T=0.1 T_K$ \cite{note} low enough so that no further increase in $\rho _{i\sigma }(\epsilon_{F})$ 
takes place as the temperature is lowered (according to physical expectations and Eq. (\ref{fsr})),
but high enough to prevent the presence of spurious spikes.
The disagreement lies below 20 \%.
The agreement improves as the parameters are moved deeper in the Kondo regime  $\epsilon_{F}-E_{1}\gg \Delta$.
Thus, while the spectral density at zero temperature is not well represented by the NCA results at $T=0$,
one can take the values at $T=0.1 T_K$ as a reasonable description of the correct $T=0$ ones.
This statement is supported by the comparison of results obtained by NCA and numerical renormalization group
for the one-level SU(2) case \cite{compa}.

\section{Kondo temperature}

From the half width at half maximum of the peak nearest to the Fermi energy
of the spectral density of the lowest level ($\rho _{1 \sigma } (\omega)$),
we have calculated the Kondo temperature of the system $T_{K}$ for several
values of $\delta$. This requires to solve the self-consistent NCA
equations up to low enough temperatures (about 0.1 $T_K$ as discussed above) 
so that the height of the peak does not
increase significantly with further lowering of the temperature \cite{note}.
Fortunately, the result is not very sensitive to the ratio $T/T_K$.

\begin{figure}
\includegraphics[width=7.5cm]{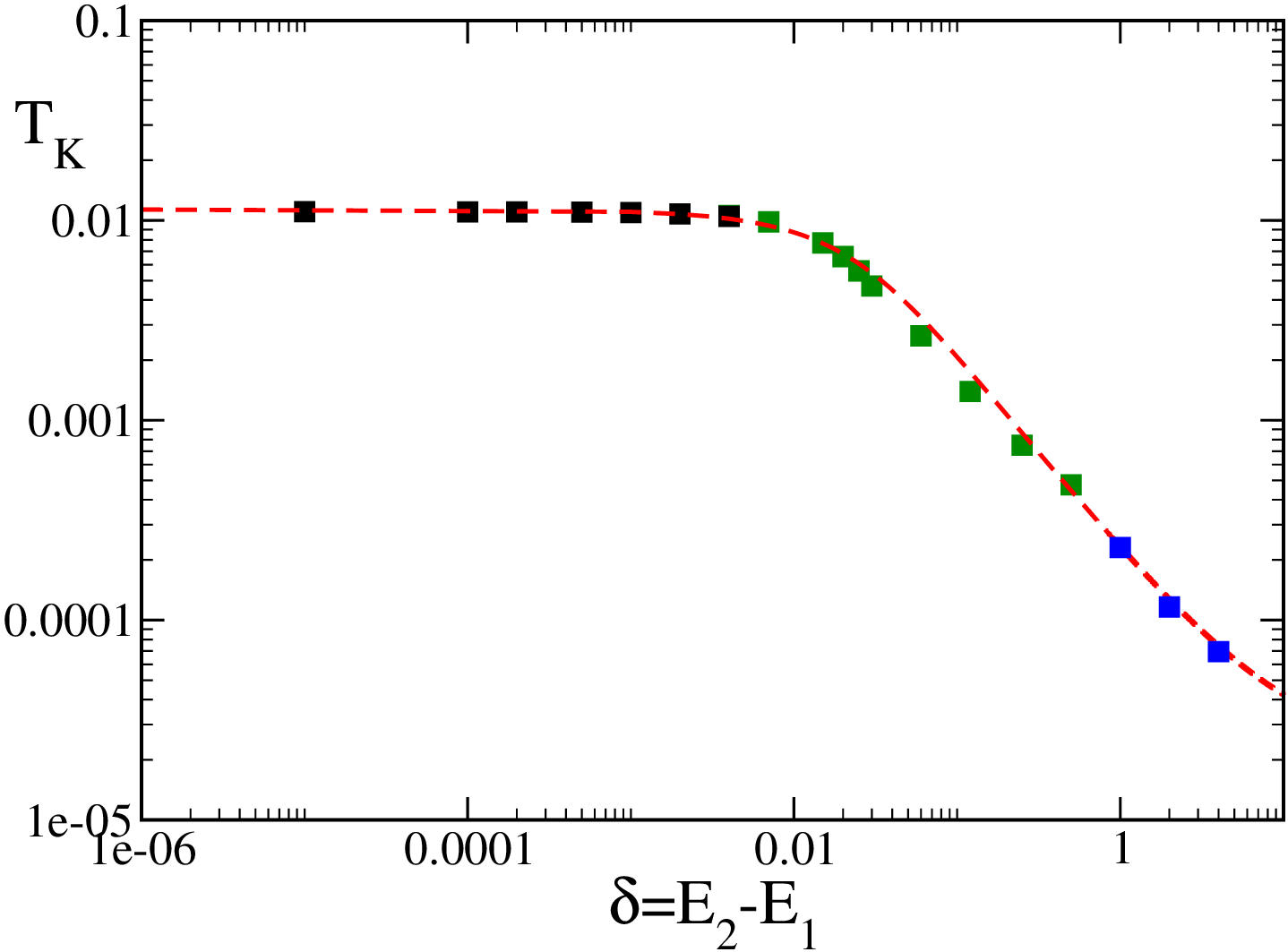}
\caption{Squares: Kondo energy scale determined by the width of the peak 
in the spectral density near the Fermi energy as a function of the splitting $\delta$. 
The temperatures used were $T=10^{-3}$, $T=10^{-4}$, and $T=10^{-5}$ depending on  $\delta$
Dashed line: corresponding variational result Eq. (\ref{tk}) multiplied by a factor 0.606.}
\label{kondo}
\end{figure}

The
results are shown in Fig. \ref{kondo}  and compared with Eq. (\ref{tk}) obtained from
a variational calculation as explained below. We see that except for an
overall multiplicative factor, the agreement is quite good, in spite of the fact
that $T_{K}$ changes by nearly two orders of magnitude.

To provide an independent estimate of  $T_{K}$, we have calculated the
stabilization energy of the following variational wave function

\begin{equation}
|\psi \rangle =\alpha |s\rangle +\sum_{ik\sigma }\beta _{ik}(|i\sigma
\rangle \langle 0|c_{ik\sigma })|s\rangle ,  \label{var}
\end{equation}%
where $|s\rangle $ is the many-body singlet state with the filled Fermi sea
of conduction electrons and the state $|0\rangle $ at the localized site, while
$\alpha$ and $\beta _{ik}$ are variational parameters.
From the resulting optimized energy $E$, we can define
the stabilization energy as $T^*_{K}=E_1-E$. 
The Kondo energy scale defined in this way becomes

\begin{equation}
T^*_{K}=\left\{ (D+\delta )D\exp \left[ \pi E_{1}/(2\Delta )\right] +\delta
^{2}/4\right\} ^{1/2}-\delta /2.   \label{tk}
\end{equation}%
This expression interpolates between the SU(4) result ($T^*_{K}=D\exp \left[
\pi E_{1}/(4\Delta )\right] $ for $\delta=0$) and the SU(2) one for one doublet only 
($T^*_{K}=D\exp \left[ \pi E_{1}/(2\Delta )\right] $ for  $\delta \rightarrow + \infty $).

\bigskip 

\section{Summary and discussion}

We have studied an impurity Anderson model containing two doublets, which
interpolates between the cases for one level with SU(4) and SU(2) symmetry, 
and is of interest for several
nanoscopic systems, using the non-crossing approximation (NCA). 

We have
shown that the NCA provides reasonable results for the equilibrium spectral
density. The values of the spectral density for both doublets agree within
20\% with the predictions of the Friedel sum rule, in spite of the fact that
it is not expected to satisfy Fermi liquid relationships at zero
temperature. 

In addition, the Kondo temperature scale $T_K$ obtained from the width of
the peak in  spectral density near the Fermi energy agrees very well with the
stabilization energy of a variational calculation, in spite of the change in
several orders of magnitude of $T_K$ when the splitting between
both doublets is changed.

The approach seems promising for studying transport properties within a
non-equilibrium formalism. Work in this direction is in progress.

\section{Acknowledgments}

One of us  (A. A. A.) is partially supported by CONICET, Argentina.  This work was partially
supported by PIP No 11220080101821 of CONICET, and PICT Nos 2006/483 and
R1776 of the ANPCyT.



\begin{thebibliography}{99}
\bibitem{note} Some technical details that allowed us to reach such low 
temperatures are briefly explained elsewhere \cite{nca2}.
\end{thebibliography}
\end{document}